\documentclass[12pt]{iopart}
\usepackage{amssymb}
\usepackage{graphicx}

\def\bef{\begin{figure}}
\def\eef{\end{figure}}

\newcommand{\be}[1]{\begin{equation}\label{#1}}  
\newcommand{\beq}{\begin{equation}}
\newcommand{\eeq}{\end{equation}}
\newcommand{\bsub}[1]{\begin{subequations}\label{#1}\begin{eqnarray}}
\newcommand{\esub}{\end{eqnarray}\end{subequations}}
\newcommand{\bea}{\begin{eqnarray}}
\newcommand{\eea}{\end{eqnarray}}
\newcommand{\bd}{\begin{displaymath}}
\newcommand{\ed}{\end{displaymath}}

\newcommand{\eq}[1]{(\ref{#1})}

\newcommand{\gev}{\,\mathrm{GeV}}

\newcommand{\alfaw}{\alpha_\mathrm{w} }
\newcommand{\alfas}{\alpha_\mathrm{s} }

\def\ARNPS{Annu.\ Rev.\ Nucl.\ Part.\ Sci.\ }

\def\IJMPit{{\it Int.\ J.\ Mod.\ Phys.\ }}

\def\MPLit{{\it Mod.\ Phys.\ Lett.\ }}

\newcommand{\bj}[9]{{#1}{#2}{#3, }{{\it #8}, }{#7 }{{#4}}{\textbf{#5} }{#6}{#9}} 
\newcommand{\bjit}[9]{{#1}{#2}{#3, }{{\it #8}, }{#7 }{{\it #4}}{\textbf{#5} }{#6}{#9}} 
\newcommand{\bjrusso}[4]{{#4 }{{#1}}{\textbf{#2} }{#3}} 
\newcommand{\bjrussoit}[4]{{#4 }{{\it #1}}{\textbf{#2} }{#3}} 

\newcommand{\equi}{\mathrm{eq}}
\newcommand{\nequi}{n_\equi}
\newcommand{\sph}{\eta_\mathrm{w}}
\newcommand{\gsph}{\Gamma_\mathrm{w} }
\newcommand{\tsph}{T_\mathrm{w} }
\newcommand{\tsphstar}{T_{\mathrm{w}}^{\star}}
\newcommand{\ncs}{{\Delta N}}

\newcommand{\sphs}{\eta_\mathrm{s} }
\newcommand{\etas}{\eta_\mathrm{s} }
\newcommand{\gs}{\Gamma_\mathrm{s} }
\newcommand{\ts}{T_\mathrm{s}}
\newcommand{\tstars}{T_{\mathrm{s}}^{\star}}

\newcommand{\gb}{\Gamma_{b}}
\newcommand{\tb}{T_{b}}
\newcommand{\etayb}{\eta_{Y_b}}
\newcommand{\gtau}{\Gamma_{\tau}}
\newcommand{\ttau}{T_{\tau}}
\newcommand{\ltau}{L_{\tau_R}}
\newcommand{\etaytau}{\eta_{Y_\tau}}

\begin{document}

\title{Sphaleron relaxation temperatures}

\author{Lu\'{\i}s Bento}

\address{Centro de F\'{\i}sica Nuclear da Universidade de Lisboa\\
Avenida Prof. Gama Pinto 2, 1649-003 Lisboa, Portugal}


\begin{abstract}
The transition of sphaleron processes from non-equilibrium to thermal equilibrium in the early Universe is examined in detail.
The relations between the damping rates and frequencies of the weak and QCD sphaleron degeneracy parameters are determined in general form and the respective relaxation temperatures are calculated in specific scenarios.
It is pointed out that the gauge coupling constants running with energy produces strong and weak sphaleron rates closer to each other at very high temperatures and makes them larger in supersymmetric models than in the standard model case.
\end{abstract}
     
\ead{lbento@cii.fc.ul.pt } 



\maketitle

\section{Introduction}\label{sectionintroduction}

Sphaleron processes are important for 
baryogenesis~\cite{Kuzmin85,Rubakov96,Dolgov92,Cohen93}.
This is obviously true in the case of electroweak baryogenesis because the weak sphalerons are the only processes that violate baryon number.
But is also the case of leptogenesis 
mechanisms~\cite{Fukugita86,Luty92,Covi96,Buchmuller00}, 
where weak sphalerons are responsible for converting part of the generated $B-L$ asymmetry into a baryon asymmetry.
In this case the effect of weak sphalerons can be reduced to a mere constraint on a linear combination $\sph$ of quark and lepton chemical potentials~\cite{Khlebnikov88,Harvey90,Khlebnikov96}
of the form $\sph=0$. 
But the constraint only applies when sphalerons are in thermal equilibrium in a range of temperatures from around $10^{12} \gev$ to the electroweak phase transition.
It is important to know well this temperature range when designing a lepto-baryogenesis mechanism, as well as all the other reactions in thermal equilibrium at every moment as the Universe cools down, because too many constraints enforce a trivial particle asymmetry result.
On the other hand, the surviving baryon asymmetry depends quantitatively on the consecutive sets of constraints applying at and after the epoch of baryogenesis or leptogenesis. 

The weak sphaleron frequency equalizes the Hubble expansion rate at a certain temperature.
However, it cannot be identified with the relaxation temperature of the sphaleron chemical potential constraint.
It is the main purpose of this paper to study the evolution of $\sph$ and evaluate its
relaxation temperature as well as for the QCD sphaleron~\cite{McLerran91,Mohapatra92} degeneracy parameter, $\sphs$.
It is also argued that the running of the weak and strong coupling constants with the energy scale leads to smaller values of the sphaleron rates at very high energies, specially in the case of strong sphalerons.
In the next section we establish the general relation between charge violation rates and sphaleron diffusion rates using standard scattering theory considerations.
In the subsequent sections we specialize to the cases of weak and strong sphalerons obtaining general results for the damping rates of the parameters $\sph$ and $\sphs$.
We calculate also the respective relaxation temperatures in the standard model case and in a typical leptogenesis scenario.
In section \ref{sectionbr} we analyze the interplay between weak sphalerons, bottom quark an tau lepton Yukawa couplings while they approach thermal equilibrium and derive the appropriate system of coupled equations.
In section \ref{sectionsupersymmetric} we extend the study to the context of supersymmetric standard model versions and in the last section summarize the conclusions.

\section{Sphaleron processes}\label{sphaleron}

The rate per unit of volume of any reaction with one or more particles 
$a_i$ in the initial state and $b_j$ in the final state,
$a_i \rightarrow b_j$,
depends on the transition amplitude $\mathcal{T}$ and particle distribution functions $f_\alpha$ as follows~\cite{Weinberg79,Kolb80,Kolb90}:
\begin{eqnarray} 	\label{g+}
	\gamma_+ = \int  d \Phi \, | \mathcal{T}|^2 \,  f_{a_i} (1 \pm f_{b_j} ) 
	\; ,													
\end{eqnarray}
where a product over $i$ and $j$ is implicit.
$d \Phi$ stands for the phase space element and  
$1 \pm f_{b_j}$ are the final state boson stimulated emission and/or fermion Pauli blocking factors.
The evolution of any charge asymmetry depends on the difference between the rates 
of $a_i \rightarrow b_j$ and its inverse reaction $b_j \rightarrow a_i$,
$\gamma_+$ and $\gamma_-$ respectively.
Using Bose-Einstein and Fermi-Dirac distribution functions and assuming that the transition amplitude is invariant under time reversal, 
one obtains to first order in the chemical potentials, 
\begin{eqnarray} 	\label{g+g-}
	\gamma_+ -\gamma_- = - \frac{1}{2} (\gamma_+ + \gamma_-) 
	\sum_\alpha \frac{\mu_\alpha}{T} \Delta N_\alpha 
	\; ,													
\end{eqnarray}
where $\Delta N_\alpha$ is $+1$ ($-1$) for each of the outgoing (incoming) particles.
An observable $Q$ that varies $\Delta Q = Q_\alpha \ncs_\alpha$ ($-\Delta Q$) 
in a single reaction 
$a_i \rightarrow b_j$ ($b_j \rightarrow a_i$) 
suffers a net violation per unit of time in a total volume $V$ equal to
${dQ}/{dt} = (\gamma_+ - \gamma_-) \,V \Delta Q$.

Sphalerons processes produce or destroy fermion particles and are characterized by a Chern-Simons number increase $\ncs$ whose sign changes with the direction of the reaction as any other charge.
Then, the violation rate of any charge $Q$ is obtained by summing over all $\ncs$ numbers as follows:
\begin{eqnarray}	\label{dqdt1}
	\frac{dQ}{dt} = - \frac{1}{2} \sum_\ncs \gamma_\ncs \, \Delta Q  
	\sum_\alpha \frac{\mu_\alpha}{T} \Delta N_\alpha \,V 
		\; .																			
\end{eqnarray}
It happens that the fermion number variations 
$\Delta N_\alpha$ are directly proportional to the Chern-Simons number variation $\ncs$, 
so that $\Delta Q/ \ncs$ and the degeneracy parameter 
\beq
\eta = \sum_\alpha \frac{\mu_\alpha}{T} \frac{\Delta N_\alpha}{\ncs} 
																				\label{eta}
\eeq
are independent of $\ncs$.
This gives 
\begin{eqnarray}
	\frac{dQ}{dt} = - \frac{1}{2} \frac{\Delta Q}{\ncs}\, \Gamma \, \eta \,V 
		\; ,																					\label{dqdt}		
\end{eqnarray}
\begin{eqnarray}
	\Gamma = \sum_{\ncs} \gamma_{\ncs} (\ncs )^2
		\; 	,																		\label{gamma1}
\end{eqnarray}
where $\Gamma$ is nothing but the sphaleron Chern-Simons number diffusion rate calculated in the literature~\cite{Khlebnikov88,Arnold97},
\begin{eqnarray}	\label{gamma2}
	\Gamma = \lim_{V, \, t \rightarrow \infty} 
	\frac{\langle (N(t) - N(0) )^2 \rangle  }{V\, t}
	\; .																			
\end{eqnarray}

\section{Weak sphalerons}\label{weak}
 
Weak sphaleron processes of Chern-Simons number variation
$\ncs$ produce fermion number violations~\cite{thooft76,Adler69} equal to
$\Delta N_{l_i} = \ncs$ per lepton doublet $l_i$
and $\Delta N_{q_i} = \ncs$ 
per quark doublet and colour state $q_i$,
where $i$ denotes the fermion generation.
Specializing the general expressions (\ref{eta}) and (\ref{dqdt}) to weak sphalerons, one derives the baryon and lepton number violation rates per generation,
\begin{eqnarray}	\label{dBqidt}
	\frac{dB_{q_i}}{dt} = 	\frac{dL_{l_i}}{dt} =	
	 - \frac{1}{2} \gsph \, \sph \,V
		\;,																				
\end{eqnarray}
where the degeneracy parameter,
\begin{eqnarray} \label{sph}
	\sph = \frac{1}{T} \sum_i (3 \mu_{q_i} + \mu_{l_i})
	\; ,														
\end{eqnarray}
is a function of the quark and lepton doublet chemical potentials, degenerate in colour and isospin during the electroweak symmetric phase.
One recovers so the known result for the baryon number violation rate~\cite{Khlebnikov88,Bochkarev87,Dine90}
(notice that $\frac{1}{2} \gsph$ represents the average between the forward ($\ncs >0$) and backward sphaleron rates).

The detailed balance equation $\sph =0$ holds 
if weak sphalerons are in thermal equilibrium, but not otherwise.
$\sph$ is a dynamical damped quantity whose relaxation time is much larger than the age of the Universe when sphalerons are not in equilibrium and much smaller when they are. 
The $\sph$ evolution depends not only on sphalerons but also on all the other reactions occurring at that moment.
The rapid reactions enforce detailed balance equations that reduce the number of independent chemical potentials.
The latter have a one-to-one relation with an equal number of independent quantum numbers that are either conserved or are violated solely by reactions not fast enough to be in full thermal equilibrium.
Without loss of generality, one can choose the independent quantum numbers as the baryon number $B$ and a set of charges, $Q_\alpha$, that are conserved by weak sphalerons:
if a given quantum number is violated by weak sphalerons there is always a linear combination with $B$, for example $B-L$, that is conserved by them. 
$\sph$ can be written in terms of $B$ and $Q_\alpha$ as
\begin{eqnarray}		\label{sphbq}
	\sph T^3 \, V = a \, B + b_\alpha \, Q_\alpha
	\; 		.																					
\end{eqnarray}
After reheating the total entropy is conserved and so is the product $T^3 V$, whenever the number of relativistic degrees of freedom is stable.
Under these conditions, and taking the usual approximate expressions, linear in the chemical potentials, of the particle anti-particle number asymmetries,
the coefficients $a$ and $b_\alpha$ are well defined constants that only depend on the kind of particles and reactions in thermal equilibrium.
They are thus model dependent but do not depend on the origin of the particle asymmetries.

Let us assume that at the temperature range of interest weak sphalerons are the only processes that violate baryon number and that all reactions besides them can be classified as either rapid or negligible slow.
The first enforce the chemical potential constraints, the latter do not contribute to flavour or any other quantum number violation rates.
In this case the charges $Q_\alpha$ are exactly conserved quantities by construction, whose values are determined by the initial conditions and particular baryogenesis or leptogenesis mechanism.
Then, one derives from equations (\ref{dBqidt}) and (\ref{sphbq}) that for 3 fermion generations,
\begin{eqnarray}		\label{dotsph}
\frac{d \sph}{dt} = - \frac{3}{2}\, a \frac{\gsph}{T^3} \,\sph
 	\; .																							
\end{eqnarray}
The sphaleron rate goes as $\gsph \sim T^4$ and in a radiation dominated Universe the Hubble expansion rate behaves as $H = 1/2t \sim T^2$.
Hence, $\tsph = \gsph  T^{-2} H^{-1}$ 
is a constant quantity. 
$\tsph$ may be naively considered as the sphaleron equilibrium temperature because $\gsph  T^{-3} = H$ at $T = \tsph$, but this is an oversimplification. 
In fact, one obtains that $\sph$ has a relaxation temperature equal to
$\tsphstar = 3/2 \, a\, \tsph$:
\begin{eqnarray}		\label{dsphdT}
\frac{d \sph}{d T^{-1}} = - \frac{3}{2}\,  a\, \tsph \, \sph
 	\; .																							
\end{eqnarray}
The parameter $\sph$ decays exponentially as $e^{-\tsphstar/T}$ and its initial value follows from equation (\ref{sphbq}) with the initial baryon number asymmetry.
As $\sph$ decays, $B$ converges to the value that annihilates the right-hand member of equation (\ref{sphbq}).

The coefficients $a$ and $b_\alpha$ are model dependent. 
They depend on the particle content and reactions in equilibrium.
The standard model interactions that enter in equilibrium before the
$10^{12} \gev$ temperature scale are the gauge interactions, implying zero gauge boson chemical potentials and fermion chemical potentials degenerate in colour and isospin,
QCD sphalerons~\cite{McLerran91,Mohapatra92}, and the
top quark Yukawa coupling ($q_t t_R^c \phi$).
They put the following constraints
on the Higgs doublet $\phi$, up and down quark isosinglets $u_i = u_R, c_R, t_R$, $d_i = d_R, s_R, b_R$, and quark doublets $q_i = q_u, q_c, q_t$ defined as having diagonal Yukawa couplings with the up quarks:
\begin{eqnarray} \label{fc1}
\sum_i (2 \mu _{q_i}-\mu _{u_i}-\mu _{d_i})  = 0
		\;,  \label{qcdinstantons} \\
\mu _{q_t}-\mu _{t_R}+\mu _{\phi} = 0
		\;,  \label{tyukawa}
\end{eqnarray}
The other constraint is the initial condition of zero hypercharge,
\beq
\sum_i (\mu_{q_i} +2 \mu_{u_i} - \mu_{d_i} - \mu_{l_i} - \mu_{e_i} ) 
+ 2 \mu_{\phi} =0
		\; ,																\label{y}
\eeq
to first order in the chemical potentials and neglecting thermal masses.
Above the $10^{12} \gev$ temperature scale
the other quark and lepton Yukawa couplings are negligible and the gauge interactions, including sphaleron processes, conserve several flavour quantum numbers, namely: 
lepton numbers $L_{e_i}$ of $e_i = e_R, \mu_R, \tau_R$; 
differences between the baryon numbers of right-handed charm or down quarks and the $u_R$ baryon number, 
$B_{c_R} - B_{u_R}$ and $B_{d_i} - B_{u_R}$;
differences between the total lepton or baryon numbers of different generations,
$L_i - L_j$ and $B_i - B_j$; $B-L$. 
While $B-L$ and $L_i - L_j$ are conserved by all standard model interactions the other quantum numbers are violated by Yukawa couplings.
The up quark is singled out because it has  the smallest quark Yukawa coupling.
As the Universe cools down and the smaller Yukawas enter in equilibrium the various flavour charges cease to be conserved and get constrained by the new detailed balance equations.

If there are non-standard interactions in equilibrium, such as lepton number violating heavy neutrino Majorana masses, more constraints apply and less independent charges subsist. 
The remaining conserved charges at any given moment, $Q_\alpha$, are determined by the particular baryogenesis or leptogenesis mechanism, but their values do not affect the $\sph$ damping rate (not even the zero hypercharge condition).
It only depends on the constraints enforced by the interactions in full equilibrium.
Assuming that only the standard model gauge interactions, strong sphalerons and top quark Yukawa coupling are in thermal equilibrium, one obtains $a = 15/2$ for the coefficient $a$ in equation (\ref{sphbq}).
It follows that the $\sph$ relaxation temperature is 
$\tsphstar = 45/4\, \tsph$, one order of magnitude larger than $\tsph$.

The sphaleron diffusion rate has been shown~\cite{Arnold97} to depend on the temperature and weak coupling constant as $\gsph \approx 26\, \alfaw^5 T^4$.
At this point we argue that the effective coupling constant runs with the energy scale and therefore varies with the temperature~\footnote{I thank M. Shaposhnikov and D. B\"odeker for discussions on this point.}.
At the $10^{12} \gev$ energy scale $\alfaw \approx 1/41$ which gives
$\gsph \approx 2 \times 10^{-7} \,T^4$,
about 5 times smaller than the rate calculated with the $M_Z$ scale value
$\alfaw \approx 1/30$.
For a Hubble rate $H = 1.66\, g_\ast^{1/2}T^2/M_\mathrm{Pl}$ with $g_\ast \approx 107$ effective number of degrees of freedom the $\tsph$ and relaxation temperature values are respectively 
$\tsph \approx 1.6 \times 10^{11} \gev$ and 
$\tsphstar \approx 1.8 \times 10^{12} \gev$.

Leptogenesis is a mechanism that has been actively investigated~\cite{Fukugita86,Luty92,Covi96,Buchmuller00}.
It is based on the addition of extra sterile neutrinos $N_a$ with very heavy Majorana masses and complex Yukawa couplings, $l_i N_a \phi$, with the standard Higgs and lepton doublets.
They cause lepton flavour and total lepton number violating processes that stay in thermal equilibrium for a certain period of time.
During that period the following constraints apply:
\beq
\mu_{l_i} + \mu_{\phi} =0
		\; .																\label{lphi}
\eeq
The quantum numbers $L_i - L_j$ and $B-L$ are then rapidly violated and consequently they are excluded from the set of charges $Q_\alpha$ in equation \eq{sphbq}.
After applying these new constraints one obtains for the coefficient $a$ the value 
$a = 261/44$ and the slightly reduced relaxation temperature 
$\tsphstar \approx 1.4 \times 10^{12} \gev$.

\section{$b_R$ and $\tau_R$ Yukawa couplings and sphalerons}\label{sectionbr}

The weak sphaleron relaxation temperature falls in the vicinity of the equilibrium temperatures of the $b_R$ and $\tau_R$ Yukawa couplings.
Our estimates of the $b_R$ and $\tau_R$ Yukawa production rates $\gb$ and $\gtau$ 
i.e., reactions where  $b_R$ and $\tau_R$ are produced in the final state, yield equilibrium temperatures equal to
$\tb = \gb T\, H^{-1} \approx 2 \times 10^{12} \gev$ and
$\ttau = \gtau T \, H^{-1} \approx 10^{12} \gev$.
This shows that one cannot neglect the violation of $Q_b \equiv B_{b_R}-B_{u_R}$ and $L_{\tau_R}$ when weak sphalerons approach thermal equilibrium.
The violation rates obtained from the basic expression (\ref{g+g-}) are given by
\bea
\frac{d Q_b}{d t}  = - 2\, \gb \, \nequi \, \etayb \, V 
		\; , 		\label{dqbdt}  \\
\frac{d \ltau }{d t} = - 2\, \gtau \, \nequi \, \etaytau \, V 
		\; , 		\label{dltaudt} 
\eea
where $\nequi = 0.90 \, T^3 / \pi^2$ denotes the ultra relativistic fermion equilibrium density and the degeneracy parameters are defined as
\bea
\etayb = \eta_{b_R} + \eta_\phi -\eta_{q_t} 
		\; , 		\label{etab}  \\
\etaytau = \eta_{\tau_R} + \eta_\phi -\eta_{l_\tau} 
		\; . 		\label{etatau} 
\eea
This goes in parallel with the baryon number violation rate,
\beq	
	\frac{d B}{d t} = - \frac{3}{2} \gsph \, \sph \,V
		\; ,																					\label{dBdt}	
\eeq
assuming that $B$ is only significantly violated by sphalerons.

The above equations have the general form,
\beq
\frac{d Q_\mu}{d t} = - \gamma_\mu \, \eta_\mu \, V
		\; ,																					\label{dqmudt}	
\eeq
where  $Q_\mu = (B, Q_b, \ltau)$ and $\eta_\mu = (\sph, \etayb, \etaytau)$.
The charges $Q_\mu$ and the set of conserved quantum numbers $Q_\alpha$ have a one-to-one relation with the independent chemical potentials left unconstrained by the detailed balance equations.
One can thus write the degeneracy parameters $\eta_\mu$
as linear combinations of $Q_\mu$ and $Q_\alpha$:
\beq
\eta_\mu \, T^3 \, V = a_\mu^\nu \, Q_\nu + b_\mu^\alpha \, Q_\alpha
		\; .																					\label{etamu}
\eeq
The differentiation against time yields the general result
\beq
\frac{d \eta_\mu}{d t} = - a_\mu^\nu \, \gamma_\nu \, T^{-3} \, \eta_\nu 
		\; .																					\label{detamudt}	
\eeq  
In a radiation dominated Universe this coupled system of equations leads to:
\bea
\frac{d \eta_\mu}{d T^{-1}} = - T_\mu^\nu \, \eta_\nu 
		\; ,																					\label{detamudT1}	\\
T_\mu^\nu =  a_\mu^\nu \, \gamma_\nu T^{-2} \, H^{-1}
		\; ,																					\label{tmunu1}
\eea  
where $T_\mu^\nu$ is the relaxation temperature matrix.

If all interactions in equilibrium at around $10^{12} \gev$ are standard model interactions,
the conserved charges $Q_\alpha$ entering in equation (\ref{etamu}) are:
$L_{e_R}$, $L_{\mu_R}$, $B_{c_R} - B_{u_R}$, $B_{s_R} - B_{u_R}$, 
$B_{d_R} - B_{u_R}$ and $B_i - B_j$, only negligibly violated by Yukawa couplings,
and the exactly conserved numbers $L_i - L_j$ and $B-L$. 
The constraints imposed by the $t_R$ Yukawa processes, strong sphalerons and gauge interactions determine the coefficients $a_\mu^\nu$ and $b_\mu^\alpha$.
In particular,
\beq
a_\mu^\nu = 6 \left(
\begin{array}{rrr}
{\frac{5}{4}}&{0}&{-\frac{1}{2}} \\
{0}&{\frac{40}{23}}&{\frac{1}{23}} \\
{-\frac{1}{6}}&{\frac{12}{23}}&{\frac{38}{23}}
\end{array}
\right)
		\; .																					\label{amunu}	
\eeq 
The estimates of the reaction rates in equations (\ref{dqbdt})-(\ref{dqmudt})
given above and at the end of the previous section produce the following relaxation temperature matrix:
\beq
T_\mu^\nu \approx \left(
\begin{array}{rrr}
2 	& 0 	& {-\frac{1}{2}} \\
{0}	& 4 	& {\frac{1}{20}} \\
{-\frac{1}{4}} & 1 & 2
\end{array}
\right)						\times 10^{12} \gev
		\; .																					\label{tmunu2}	
\eeq  
The elements 11 of these two matrices coincide with the values of the coefficient $a$ in equation (\ref{sphbq}) and the sphaleron relaxation temperature calculated in the previous section, where the $b_R$ and $\tau_R$ Yukawa couplings were neglected.
Now, $\eta_\mu = \sph$, $\etayb, \etaytau$ obey the coupled system of equations (\ref{detamudT1}).
It can be integrated in closed form:
diagonalizing $T_\mu^\nu$ with a non-orthogonal matrix $V$, the general solution is, in matrix notation,
$\eta = V e^{T^\star (T_0^{-1}-T^{-1})} V^{-1} \eta_0$,
where $T^\star$ is the diagonalized relaxation temperature.
The $\eta_\mu$ parameters oscillate into each other as they decay but the outcome of this complex evolution is clear:
weak sphalerons, $b_R$ and $\tau_R$ Yukawa couplings enter in equilibrium at the same time with relaxation temperatures in the range 
$(2 - 4)\times 10^{12} \gev$.

This exercise can be repeated for periods characteristic of leptogenesis in which lepton number is rapidly violated by heavy neutrino Majorana masses.
Then, the additional constraints \eq{lphi} are enforced and the conserved charges $Q_\alpha$ in equation (\ref{etamu}) reduce to:
$L_{e_R}$, $L_{\mu_R}$, $B_{c_R} - B_{u_R}$, $B_{s_R} - B_{u_R}$, 
$B_{d_R} - B_{u_R}$ and $B_i - B_j$. 
Now, the results for the coefficients $a_\mu^\nu$ and relaxation temperature matrix are
\beq
a_\mu^\nu = \frac{3}{22} \left(
\begin{array}{rrr}
\frac{87}{2}  & -36  & -21 \\
			-4			&  64  &  8  \\
			-7			&  24  &  58
\end{array}
\right)
		\; ,																					\label{amunu2}	
\eeq 
\beq
T_\mu^\nu \approx \left(
\begin{array}{rrr}
	\frac{3}{2}			& 	-	2			&  	- \frac{1}{2} 				\\
	- \frac{1}{8}		&    3			&			\frac{1}{5}			 		\\
	- \frac{1}{5}		&    1			& 		\frac{3}{2}
\end{array}
\right)
			\times 10^{12} \gev
		\; ,																					\label{tmunu3}	
\eeq 
in the same range of temperatures as the pure standard model case.

\section{QCD sphalerons}\label{qcd}

The value of the QCD sphaleron diffusion rate, $\gs$, is still uncertain but there are indications~\cite{Moore97,Moore00} 
that it scales with the weak sphaleron rate as 
$\gs \sim 12 \, \alfas^5 /\alfaw^5 \gsph$.
At high energies the strong and weak coupling constants converge to each other: 
$\alfas \approx 1/37$ and $\alfaw \approx 1/42$ at $10^{13} \gev$, which gives 
$\gs  \sim 4 \times  10^{-6}\, T^4$, only 20 times larger than $\gsph$.
The reference temperature in a radiation dominated Universe, 
$\ts = \gs T^{-2} H^{-1}$, takes the value 
$\ts \sim 3 \times 10^{12} \gev$.
Again, this is not the relaxation temperature, $\tstars$, of the strong sphaleron degeneracy parameter,
\begin{eqnarray}
	\etas = \frac{1}{T} \sum_i (2 \mu_{q_i} - \mu_{u_i} -\mu_{d_i} )
	\; .															\label{etas}
\end{eqnarray}
In order to study the $\etas$ dynamical evolution one needs the rates of chiral charge violation induced by QCD sphalerons.
A sphaleron with a Chern-Simons number increase $\ncs$ induces a chiral charge variation 
$\Delta Q_{5} = 2 \ncs$ for each of the 6 Dirac quarks~\cite{thooft76,Adler69,McLerran91,Mohapatra92}, 
defining chiral charge as $+1$ for left-handed and $-1$ for right-handed quarks,
which means a quark number increase 
$\Delta N_\alpha = \pm \ncs/3$
per colour and chiral state.
Employing the general result of equations (\ref{eta}), (\ref{dqdt}) and  (\ref{gamma1}), the violation rates of left-handed ($q_L$) and right-handed ($q_R$) quark number asymmetries are
\begin{eqnarray}
\frac{dN_{q_L}}{dt} = - \frac{dN_{q_R}}{dt} = - \frac{1}{2}\, \gs \, \etas \, V
 	\; 			,																					\label{dnqldt}
\end{eqnarray}
per isospin state (summing over colours).
This is in agreement with ref.~\cite{Moore97}
(again, $\frac{1}{2} \gs$ represents the average between forward and backward sphaleron rates).

Neglecting thermal mass corrections, 
$\etas$ is directly related with the total chiral charge as
$Q_5 = \frac{1}{2} \etas T^3 V$,
but that is not enough to determine the $\etas$ evolution from equation (\ref{dnqldt}) 
because the third generation chiral charge is rapidly violated by the top quark Yukawa coupling.
It is more convenient to express $\etas$ in terms of the right-handed up quark baryon number
$B_{u_R} = N_{u_R}/3$, 
for example ($u_R$ is the quark with smallest Yukawa coupling), 
and a set of independent quantum numbers $Q_\alpha$ conserved by strong sphalerons: 
the baryon number $B$ and other charges conserved by both strong and weak sphalerons:
\begin{eqnarray}
\etas T^3\, V = - a\, B_{u_R} + b_\alpha \, Q_\alpha
 	\; .																									\label{etasbuq}
\end{eqnarray}
The coefficients $a$, $b_\alpha$ and the set of charges $Q_\alpha$ depend on which interactions are in thermal equilibrium.
However, if all processes that are not in full thermal equilibrium including weak sphalerons can be considered negligibly slow, the charges $Q_\alpha$ are conserved in  first approximation and only $B_{u_R}$ is violated in the right-hand side of equation (\ref{etasbuq}).
One derives
\begin{eqnarray}
\frac{d  \etas}{dt} = - \frac{a}{6}\, \frac{\gs}{T^3}\, \etas
 	\; ,																									\label{dsdt}
\end{eqnarray}																									
which implies a relaxation temperature equal to $\tstars = a\, \ts/6$ 
in a radiation dominated Universe:
\begin{eqnarray}
\frac{d  \etas}{d T^{-1}} = - \frac{a}{6} \, \ts \, \etas
 	\;   .																									\label{detasdT}
\end{eqnarray}

If no interactions beyond standard model are in thermal equilibrium, the chemical potential constraints are determined by the gauge interactions, top quark Yukawa coupling and the condition $Y = 0$
(equations (\ref{tyukawa}) and (\ref{y})).
The approximately or exactly conserved charges $Q_\alpha$ in equation (\ref{etasbuq}) are 
$B$, $B-L$, $L_{e_i}$, $B_{c_R} - B_{u_R}$, $B_{d_i} - B_{u_R}$, $L_i - L_j$ and $B_i - B_j$. 
In that case one calculates $a = 184/3$ and the consequent relaxation temperature is 
$\tstars = 10.2\, \ts \sim 3 \times 10^{13} \gev$.
It compares with the value obtained if we had ignored the top quark Yukawa coupling, $\tstars = 12 \, \ts$.

$\etas$ decays as $e^{-\tstars/T}$ and its initial value is obtained from equation (\ref{etasbuq}) with the initial $B_{u_R}$ value.
As $\etas$ decays, $B_{u_R}$ converges to the value that annihilates the right-hand side of equation (\ref{etasbuq}).
When the weak sphalerons enter in equilibrium at around $2 \times 10^{13} \gev$,
$\etas$ is already suppressed by $e^{-\tstars/T} \sim e^{-15}$,
which justifies the approximation of taking the relaxed value $\etas =0$ as we did in the two previous sections.

\section{Supersymmetric models}\label{sectionsupersymmetric}

Baryogenesis and leptogenesis have been also studied in the context of supersymmetric theories.
The supersymmetry breaking terms decouple at a certain temperature 
$T_\mathrm{ss} \sim 10^7 \gev$
and supersymmetry becomes a good symmetry above $T_\mathrm{ss}$.
As a result, other global symmetries become effective at high temperatures namely,
in the case of the minimal supersymmetric standard model (MSSM), 
the Peccei-Quinn, U(1)$_A$, and U(1)$_R$ symmetries.
It was shown~\cite{Ibanez92} that in addition to $B-L$ another linear combination, $R'$, of $A$, $R$, $B$ and $L$ is anomaly free under the gauge groups SU(3) and SU(2).
Thus, $R'$ and $B-L$ are unconstrained conserved quantum numbers even if all reactions including sphaleron processes are in thermal equilibrium.
Then, a net baryon number asymmetry requires non-zero values of either $R'$ or $B-L$ but not necessarily both of them, in contrast with the non-supersymmetric case where 
a non-zero $B-L$ is needed.

But at very high temperatures there are in fact other approximate symmetries because
some of the lepton and quark Yukawa couplings are thermally decoupled as discussed in the previous sections.
In addition, as we prove in the following, supersymmetry adds two and not just one SU(3) and SU(2) anomaly free conserved quantum numbers.

\begin{table}
	\caption{\label{Susy}Quantum numbers and anomalies in the supersymmetric standard model.}
\lineup
\begin{indented}
\item[]
\begin{tabular}{@{}lllllllllllll}
\br
 & $\tilde{A}$ & $\0 \tilde{H}_1$ & $\; \tilde{H}_2$ &  $\m {q}$ & $\0 {u}^C $ & $\0 {d}^C $ 
 & $\m {l}$ & $\0 {e}^C$ & $\0 {N}$			 & $\0 a_3$ & $\0 a_2$ & $\0 a_1$		 \\
\mr
$Y$ & $0$ & $-1$ & $\0 1$ &				$\m \frac{1}{3}$	& $-\frac{4}{3}$ & $\m \frac{2}{3}$ 
 & $-1$ & $\m 2$ & $\m 0$			 		 & $\m 0$ & $\m 0$ & $\m  0$		 \\
$B$ & $0$ & $\m 0$ & $\0 0$ &			$\m \frac{1}{3}$	& $-\frac{1}{3}$ & $-\frac{1}{3}$ 
 & $\m 0$ & $\m 0$ & $\m 0$			     & $\m 0$ & $\m 3$ & $ -3$		 \\
$L$ & $0$ & $\m 0$ & $\0 0$ &  		$\m 0$ &					$\m 0$					 & $\m 0$ 
 & $\m 1$ & $-1$ & $-1$			   & $\m 0$ & $\m 3$ & $ -3$		 \\
$A$ & $0$ & $\m 1$ & $\0 0$ &  		$\m 0$ & $\m 0$ & $-1$ 
 & $\m 0$ & $-1$ & $\m 0$			     & $-3$ & $\m 1$ & $ -7$		 \\ 
$R$ & $1$ & $\m 0$ & $\0 0$ &  		$-\frac{1}{2}$ & $-\frac{1}{2}$ & $-\frac{1}{2}$ 
 & $-\frac{1}{2}$ & $-\frac{1}{2}$ & $-\frac{1}{2}$			   & $\m 0$ & $-2$ & $-10 $		 \\
$R_M$ &$3$ & $\m 0$ & $\0 0$ &  			 $-\frac{1}{3}$ & $-\frac{8}{3}$ & $-\frac{8}{3}$ 
 & $-3$ & $\m 0$ & $\m 0$			   & $\m 0$ & $\m 0$ & $-36$		 \\
\br			
		\end{tabular}
\end{indented}
\end{table}

The exact supersymmetric part of MSSM comprises the SU(3)$\times$SU(2)$\times$U(1)$_Y$ gauge interactions and the Yukawa couplings specified by the superpotential
\beq
W = \lambda_u H_2 \, q\, u^C + \lambda_d H_1 \, q\, d^C + \lambda_e H_1 \, l\, e^C 
				\; ,			\label{W}   
\eeq
where the flavour indices are omitted.
In the case of supersymmetric 
leptogenesis~\cite{Covi96,Murayama93,Campbell93,Plumacher98,Lazarides98}
one adds heavy sterile neutrinos $N_a$ with Majorana masses $M_a$ and the superpotential
\beq
W_N = \frac{1}{2} M N N + \lambda_N H_2 \, l\, N  
				\; .			\label{WN}   
\eeq
The MSSM interactions conserve $B$ and $L$ and two other quantum numbers, $A$ and $R$, of the global Peccei-Quinn and U(1)$_R$ symmetries.
The particle quantum numbers are specified in table \ref{Susy}
($R$ as defined in~\cite{Lazarides98,Hall91}).
$\tilde{H}_1$ and $\tilde{H}_2$ are the Higgsino fields and
$\tilde{A}$ stand for the left-handed components of the gauginos $\tilde{W}$, $\tilde{B}$ and $\tilde{g}$.

The $R$ charges of the Higgs fields $H_1$, $H_2$ and scalar superpartners of $q$, $u^C$, $d^C$, $l$, $e^C$, $N$ exceed the respective fermion superpartner charges by 1 unit so that the superpotential carries 2 units.
The gauge bosons $W$, $B$ and $g$ have zero $A$ and $R$ charges.
$R_M$ is the linear combination
\beq
 R_M = 3R + \frac{7}{2}B - \frac{3}{2} L
\; ,				\label{RM}
\eeq
defined in such a way as to be non-anomalous under SU(3) and SU(2), and conserved by sterile neutrino Majorana masses and Yukawa couplings, if applicable.
The $R_M$ charges of the Higgs and matter scalar particles exceed their fermion superpartner charges by 3 units and the superpotential carries 6 $R_M$ units.

The last three columns of table \ref{Susy} represent the anomalies under the standard model gauge groups SU(3), SU(2) and U(1)$_Y$, respectively, normalized to coincide with the quantum number variations per unit of Chern-Simons number, $\ncs$.
For example, the baryon number varies as $\Delta B = 3 \ncs$ under SU(2) weak sphalerons while $\Delta A = \ncs$.
The Higgsinos and gauginos contribute to the anomalies as follows:
the Higgsino fermion numbers vary as 
$\ncs_{\tilde{H_1}} = \ncs_{\tilde{H_2}} = \ncs$
under weak sphalerons and the Wino left-handed components 
as $\ncs_{\tilde{W}} = 4 \ncs$.
The gluino left-handed components vary as $\ncs_{\tilde{g}} = 6 \ncs$ under SU(3) QCD sphalerons.
As a result the chemical potential constraints enforced by weak and strong sphaleron processes when they are in thermal equilibrium are modified with respect to the standard model.
Equations (\ref{sph}) and (\ref{etas}) are replaced with
\begin{eqnarray}
	\sph = \sum_i (3 \eta_{q_i} + \eta_{l_i} ) + 
	\eta_{\tilde{H_1}} + \eta_{\tilde{H_2}} + 4 \eta_{\tilde{W}}
	\; ,														 \label{sphsusy}		\\
	\etas = \sum_i (2 \eta_{q_i} - \eta_{u_i} -\eta_{d_i} ) +
	6 \eta_{\tilde{g}}
	\; .															\label{etassusy}
\end{eqnarray}

In order to determine the $\sph$ and $\etas$ damping rates it is necessary to identify the quantum numbers conserved by the respective sphaleron processes.
Besides the hypercharge and $B-L$ there are flavour dependent quantum numbers that are conserved by both weak and strong sphalerons.
Within the standard model the partial lepton numbers of the lepton iso-singlets, $L_{e_i}$, the differences between the right-handed charm or down quarks baryon numbers and the $u_R$ baryon number, 
$B_{c_R} - B_{u_R}$ and $B_{d_i} - B_{u_R}$,
the differences between total lepton and baryon numbers of different generations,
$L_i - L_j$ and $B_i - B_j$ 
are only violated by Yukawa couplings.
At high enough temperatures these are out of equilibrium and the respective flavour quantum numbers are conserved to a very good approximation.
This remains true in the supersymmetric version of the standard model simply by extending the partial lepton and baryon numbers to the s-leptons and s-quarks scalar superpartners.
But there are two more conserved charges.
One is $R_M$ as discussed above.
The Peccei-Quinn charge $A$, on the contrary, is not conserved by strong sphalerons, nor any combination of $A$, $B$ and $L$.
However, a combination like
\beq
A_u = A - \frac{1}{3} B - 9 B_{u_R}
					\label{Au}
\eeq
is not anomalous under SU(3) and SU(2) and is conserved as long as the up quark Yukawa coupling remains out of equilibrium, which happens at temperatures above $10^8 \gev$.
Thus, supersymmetry adds two new conserved charges, $A_u$ and $R_M$, at very high temperatures.

Another significant consequence of supersymmetry is a different evolution of the gauge coupling constants which reflects on the sphaleron rates that scale with the fifth powers of $\alfas$ and $\alfaw$.
At $10^{12} \gev$, $\alfaw \approx 1/26$ instead of the standard model value 
$\alfaw \approx 1/41$.
It makes the weak sphaleron diffusion rate 10 times larger than in the non-supersymmetric case: 
$\gsph \approx 25\, \alfaw^5 \, T^4 \approx 2 \times 10^{-6} \,T^4$.
A similar increase occurs for QCD sphalerons.
At $10^{14} \gev$, $\alfas \approx 1/21.5$ 
yields a diffusion rate, 
$\gs \sim 300\, \alfas^5  \,T^4 \approx 6 \times 10^{-5} \,T^4$,
15 times larger than in standard model case ($\alfas \approx 1/39$).

\subsection{QCD sphalerons}\label{subqcd}

The determination of the strong sphaleron degeneracy parameter $\etas$ proceeds like in the standard model case.
One writes down the chemical potentials in terms of a set of independent global quantum numbers that are conserved by all significant processes with the possible exception of QCD sphalerons and obtains a relation like (\ref{etasbuq}) for the quantity defined in
equation (\ref{etassusy}):
\beq
\etas T^3\, V = - a\, B_{u_R} +  b_\alpha \, Q_\alpha 
 	\; .																									\label{etasbuqsusy}
\eeq
$Q_\alpha$ are all charges conserved in a temperature range around $10^{14} \gev$.
If the MSSM interactions are the only ones in thermal equilibrium, 
the charges $Q_\alpha$ are the baryon number $B$,
only marginally violated by weak sphalerons,
and the anomaly free
$B-L$, $R_M$, $A_u$, $L_{e_i}$, $B_{c_R} - B_{u_R}$, $B_{d_i} - B_{u_R}$,
$L_i - L_j$ and $B_i - B_j$.
If other non-standard interactions are in equilibrium new constraints apply and the number of conserved charges is reduced.
In any case the variation of $\etas T^3\, V$ is always the one of $B_{u_R}$ times the factor $-a$.
Equation (\ref{dsdt}) and (\ref{detasdT}) still hold and the relaxation temperature is given by
$\tstars = a\, \ts /6 $.
What are different are the coefficients $a$ and $b_\alpha$, which depend on the actual degrees of freedom and interactions in equilibrium.

The gauge and gaugino interactions put constraints of the general form
\beq
\mu_{\tilde{f}_L} = \mu_{f_L} + \mu_{\tilde{A}}
	\;	,				\label{tildef}
\eeq
where $f_L$ stands for any of the left-handed fermion particles
$q_i$, $u_i^C$, $d_i^C$, $l_i$, $e_i^C$, $\tilde{H}_1$, $\tilde{H}_2$
and ${\tilde{f}_L}$ is the respective scalar superpartner. 
$\mu_{\tilde{A}} = \mu_{\tilde{g}} = \mu_{\tilde{W}} = \mu_{\tilde{B}}$
is the common (left-handed) gaugino chemical potential whereas the gauge bosons have zero chemical potentials.
The other constraints are enforced by the top quark Yukawa coupling,
\beq
\mu_{q_t} - \mu_{t_R} + \mu_{H_2} = 0
	\; ,			\label{tyukawa2}
\eeq
and the condition of zero total hypercharge. 
In the usual linear approximation this amounts to:
\bea   
  \sum_i (\mu_{q_i} +2 \mu_{u_i} - \mu_{d_i} - \mu_{l_i} - \mu_{e_i} ) 
+ \mu_{\tilde{H_2}} - \mu_{\tilde{H_1}} 
\nonumber \\  
+ 2 \sum_i (\mu_{\tilde{q}_i} +2 \mu_{\tilde{u}_i} - \mu_{\tilde{d}_i} - 
\mu_{\tilde{l}_i} - \mu_{\tilde{e}_i} ) + 2 (\mu_{{H_2}} - \mu_{{H_1}}) =0
		\; .							 \label{ysusy}
\eea
The constraint equations can be used to express all chemical potentials in terms of 
$B_{u_R}$ and the conserved quantities $Q_\alpha$ of equation (\ref{etasbuqsusy}).
If no interactions beyond the MSSM are in equilibrium one obtains the coefficient
$a = 9440/429$ and the relaxation temperature $\tstars = 3.67\, \ts$.
The temperature $\ts = \gs T^{-2} H^{-1} $
calculated with
$\gs \sim 300\, \alfas^5  \,T^4$,  $\alfas \approx 1/21.5$,
and an effective number of degrees of freedom $g_\ast \approx 229$
($H = 1.66\, g_\ast^{1/2}T^2/M_\mathrm{Pl}$)
amounts to $\ts \sim 3 \times 10^{13} \gev$.
It makes the relaxation temperature,
$\tstars \sim 10^{14} \gev$,
4 times larger than in the non-supersymmetric case.

\subsection{Weak sphalerons}\label{subweak}

When the weak sphalerons approach thermal equilibrium the strong sphaleron processes are already very rapid and the respective degeneracy parameter is damped down by several orders of magnitude.
One applies then the constraint $\etas =0$ on the expression (\ref{etassusy}) and the constraints (\ref{tildef}) and (\ref{tyukawa2}).
The particle chemical potentials and the weak sphaleron degeneracy parameter $\sph$ of equation (\ref{sphsusy}) can be written in terms of the baryon number and a set of quantum numbers $Q_\alpha$ conserved by both weak and strong sphalerons:
\beq
\sph T^3\, V = a\, B +  b_\alpha \, Q_\alpha 
 	\; .																									\label{sphbqsusy}
\eeq
If no other than MSSM interactions are in thermal equilibrium the $Q_\alpha$ charges are
$B-L$, $R_M$, $A_u$, $L_{e_i}$, $B_{c_R} - B_{u_R}$, $B_{d_i} - B_{u_R}$,
$L_i - L_j$ and $B_i - B_j$.
As a first estimate of the $\sph$ relaxation temperature one ignores the violation of
$L_{\tau_R}$ and $B_{b_R} - B_{u_R}$ by the tau and quark Yukawa couplings.
Then, the time derivative of $\sph T^3\, V $ is just $a\, dB/dt$ and the results 
(\ref{dotsph}) and (\ref{dsphdT}) follow from there.
We obtain $a = 8179/4720$ and the relaxation temperature 
$\tsphstar = \frac{3}{2} a\, \tsph = 2.60\, \tsph$.
Taking a diffusion rate $\gsph \approx 26\, \alfaw^5 \, T^4$ with $\alfaw \approx 1/26$,
one gets $\tsph = \gsph T^{-2} H^{-1} \approx 10^{12} \gev$ and
$\tsphstar \approx 3 \times 10^{12} \gev$.
Although $\tsph$ is 7 times larger than in the standard model, the relaxation temperature
$\tsphstar$ is just less than 2 times larger.
It remains so in the vicinity of the $b_R$ and $\tau_R$ Yukawa coupling relaxation temperatures causing the same phenomenon discussed in section \ref{sectionbr} of damped oscillations of $\sph$ and the other two degeneracy parameters.

\section{Conclusions}\label{conclusions}

Weak and strong sphaleron processes enforce chemical potential constraints at high temperatures but the respective relaxation temperatures are not the ones where the sphaleron frequencies $\Gamma\, T^{-3}$ coincide with the Hubble expansion rate.
They are larger than that for both kind of sphalerons but the exact factors depend on the set of reactions in thermal equilibrium at the relevant epoch: they are 10 times larger in the case of the standard model and around 3 in the case of supersymmetric models.
These numbers do not depend on the origin of the asymmetries i.e., on the particular lepto-baryogenesis mechanism, except if exotic particles or interactions are in thermal equilibrium at the time considered.

One aspect that has been overlooked so far concerns the evaluation of the sphaleron diffusion rates themselves.
The $\alfaw$ and $\alfas$ gauge coupling constants run with the energy scale and this causes the weak and strong sphaleron rates to approach each other at very high temperatures and to be model dependent on the other hand.
The weak sphaleron rate is about 10 times larger in supersymmetric model extensions than in the standard model while the strong sphaleron rate is enhanced by a factor of 15. 
Our calculations yield the relaxation temperature values of 
$2 \times 10^{12} \gev$ and $3 \times 10^{13} \gev$ for weak and strong sphalerons, respectively, in the standard model case and 
$3 \times 10^{12} \gev$ and $ 10^{14} \gev$ in the supersymmetric version.

Weak sphalerons enter in equilibrium in the same period as the bottom and tau Yukawa couplings.
We shown that the respective degeneracy parameters obey a coupled system of equations and oscillate into each other as they are damped away.

\ack{I thank D.~B\"odeker, G.~Moore and M.~Shaposhnikov for discussions.
This work was partially supported by the FCT grant CERN/FNU/43666/2001.}





\Bibliography{99}

\bibitem{Kuzmin85}  
\bj{Kuzmin~V~A, }{Rubakov~V}{ and Shaposhnikov~M~E}{\PL}{155B}{36}{1985}
{On the anomalous electroweak baryon-number non-conservation in the early universe}{}

\bibitem{Rubakov96}
\bjit{Rubakov~V}{}{ and Shaposhnikov M~E}{Phys. Usp.\ }{39}{461}{1996}
{Electroweak baryon number non-conservation in the early universe and in high energy collisions}{ [hep-ph/9603208]}
(\bjrussoit{Usp. Fiz. Nauk }{166}{493}{1996})

\bibitem{Dolgov92}
\bjit{Dolgov A~D}{}{}{Phys. Rep. }{222}{309}{1992}
{Non-GUT baryogenesis}{}
\nonum \bjit{Riotto~A}{}{ and Trodden~M}
{Annu.\ Rev.\ Nucl.\ Part.\ Sci.\ }{49}{35}{1999}
{Recent progress in baryogenesis}{ [hep-ph/9901362]}
\nonum \bjit{}{}{Riotto~A}{Summer School in High Energy Physics and Cosmology (29 June - 17 July Trieste)}{}{}{1998}
{Theories of baryogenesis}{ [hep-ph/9807454]}
\nonum {Dine~M}{}{ and Kusenko~A,}
{\it The origin of the matter-antimatter asymmetry}{ [hep-ph/0303065]}

\bibitem{Cohen93}
\bjit{Cohen A~G,}{ Kaplan D~B}{ and Nelson A~E}{\ARNPS}{43}{27}{1993}
{Progress in electroweak baryogenesis}{ [hep-ph/9302210]}

\bibitem{Fukugita86} 
\bj{Fukugita~M}{}{ and Yanagida~T}{\PL B }{174}{45}{1986}
{Baryogenesis without grand unification}{}
\nonum \bj{Fukugita~M}{}{ and Yanagida~T}{\PR D }{42}{1285}{1990}
{Sphaleron induced baryon number nonconservation and a constraint on Majorana neutrino masses}{}

\bibitem{Luty92}
\bj{Luty M~A}{}{}{\PR D }{45}{455}{1992}
{Baryogenesis via leptogenesis}{}

\bibitem{Covi96}
\bj{Covi L,}{ Roulet E}{ and Vissani F}{\PL B }{384}{169}{1996}
{CP violating decays in leptogenesis scenarios}{ [hep-ph/9605319]}

\bibitem{Buchmuller00}
\bj{Buchm\"{u}ller~W}{}{ and Pl\"{u}macher~M}{\IJMPit A }{15}{5047}{2000}
{Neutrino masses and the baryon asymmetry}{ [hep-ph/0007176]}

\bibitem{Khlebnikov88}  
\bj{Khlebnikov S~Yu}{}{ and Shaposhnikov M~E}{\NP}{B308}{885}{1988}
{The statistical theory of anomalous fermion number non-conservation}{}

\bibitem{Harvey90}
\bj{Harvey J~A}{}{ and Turner M~S}{\PR D }{42}{3344}{1990}
{Cosmological baryon and lepton number in the presence of electroweak fermion number violation.}{}

\bibitem{Khlebnikov96}
\bj{Khlebnikov S~Yu}{}{ and Shaposhnikov M~E}{\PL B }{387}{817}{1996}
{Melting of the Higgs vacuum: conserved numbers at high temperature}{ [hep-ph/9607386]}
\nonum \bj{Laine~M}{}{ and Shaposhnikov~M}{\PR D }{61}{117302}{2000}
{A remark on sphaleron erasure of baryon asymmetry}{ [hep-ph/9911473]}
\bibitem{McLerran91}
\bj{McLerran~L, }{Mottola~E}{ and Shaposhnikov M~E}{\PR D }{43}{2027}{1991}
{Sphalerons and axion dynamics in high-temperature QCD}{}

\bibitem{Mohapatra92}
\bj{Mohapatra R~N}{}{ and Zhang~X}{\PR D }{45}{2699}{1992}
{QCD sphalerons at high temperature and baryogenesis at the electroweak scale}{}

\bibitem{Weinberg79}
\bj{Weinberg~S}{}{}{\PRL}{42}{850}{1979}{Cosmological production of baryons}{}

\bibitem{Kolb80}
\bj{Kolb E~W}{}{ and Wolfram~S}{\NP}{B172}{224}{1980}
{Baryon number generation in the early universe}{}; \bjrusso {\NP}{B195}{542 (E)}{1982}

\bibitem{Kolb90}
\bj{Kolb E~W}{}{ and Turner M~S}{(Addison-Wesley, Reading, MA)}{}{}{1990}
{The early universe}{}

\bibitem{Arnold97}
\bj{Arnold~P, }{Son~D}{ and Yaffe L~G}{\PR D }{55}{6264}{1997}
{The hot baryon violation rate is $O(\alfaw^5 T^4)$}{ [hep-ph/9609481]}
\nonum  \bj{B\"odeker~D}{}{}{\PL B }{426}{351}{1998}
{Effective dynamics of soft non-abelian gauge fields at finite temperature}
{ [hep-ph/9801430]}
\nonum \bj{B\"odeker~D, }{Moore G~D}{ and Rummukainen~K}
{\PR D }{61}{056003}{2000}
{Chern-Simons number diffusion and hard thermal loops on the lattice}
{ [hep-ph/9907545]}
\nonum \bj{Moore G~D}{}{}{\PR D }{62}{085011}{2000}
{Sphaleron rate in the symmetric electroweak phase}{ [hep-ph/0001216]}

\bibitem{thooft76}
\bj{'t Hooft~G}{}{}{\PRL}{37}{8}{1976}
{Symmetry breaking through Bell-Jackiw anomalies}{}

\bibitem{Adler69}
\bj{Adler S~L}{}{}{\PR}{177}{2426}{1969}
{Axial vector vertex in spinor electrodynamics}{}
\nonum \bj{Bell J~S}{}{ and Jackiw~R}{\NC}{60A}{47}{1969}
{A PCAC puzzle: $\pi^0 \rightarrow \gamma \gamma$ in the sigma model}{}

\bibitem{Bochkarev87}
\bj{Bochkarev A~I}{}{ and Shaposhnikov M~E}{\MPLit A }{2}{417}{1987}
{Electroweak production of baryon asymmetry and upper bounds on the Higgs and top masses}{}

\bibitem{Dine90}
\bj{Dine~M}{}{ \it et al}{\NP}{B342}{381}{1990}
{Baryon number violation at high temperature in the standard model}{}

\bibitem{Moore97}
\bj{Moore G~D}{}{}{\PL B }{412}{359}{1997}
{Computing the strong sphaleron rate}{ [hep-ph/9705248]}

\bibitem{Moore00}
\bjit{Moore G~D}{}{}
{Strong and Electroweak Matter 2000 (Marseille, 14-17 June)}{}{}{2000}
{Do we understand the sphaleron rate?}{ [hep-ph/0009161]}

\bibitem{Ibanez92}
\bj{Ib\'a\~nez L~E}{}{ and Quevedo F}{\PL B }{283}{261}{1992}
{Supersymmetry protects the primordial baryon asymmetry}{ [hep-ph/9204205]}

\bibitem{Murayama93}
\bj{Murayama H,}{ Suzuki H}{ and Yanagida T}{\PRL }{70}{1912}{1993}
{Chaotic inflation and baryogenesis by right-handed sneutrinos}{}

\bibitem{Campbell93}
\bj{Campbell B~A,}{ Davidson S}{ and Olive K~A}{\NP }{399}{111}{1993}
{Inflation, neutrino baryogenesis, and (s)neutrino-induced baryogenesis}
{ [hep-ph/9302223]}

\bibitem{Plumacher98}
\bj{Pl\"{u}macher~M}{}{}{\NP B }{530}{207}{1998}
{Baryon asymmetry, neutrino mixing and supersymmetric SO(10) unification}
{ [hep-ph/9704231]}

\bibitem{Lazarides98}
\bj{Lazarides G}{}{ and Shafi Q}{\PR D }{58}{071702}{1998}
{R-symmetry in the minimal supersymmetric standard model and beyond with several consequences}
{ [hep-ph/9803397]}

\bibitem{Hall91}
\bj{Hall L~J}{}{ and Randall L}{\NP }{B352}{289}{1991}
{U(1)$_R$ symmetric supersymmetry}{ }

\end{thebibliography}
\end{document}